\documentclass[12pt,a4paper,reqno]{amsart}
\usepackage{graphicx} 
\usepackage{amsmath,amssymb,amsthm,array}
\usepackage{amsfonts, amsmath, amsthm, amssymb} 

\begin{document}
\title[Kdv-Burgers Equation in a Layered Medium]{Modelling Solutions to the Kdv-Burgers Equation in the Case of\\ Non-homogeneous Dissipative Media}

\author{Alexey Samokhin}\vspace{6pt}

\address{Dept. of Math., Moscow State Technical University of Civil Aviation}\vspace{6pt}

\email{ samohinalexey@gmail.com}\vspace{6pt}

\begin{abstract}
We study the behavior of the soliton  which, while moving in non-dissipative medium encounters a barrier with finite dissipation. The modelling included the case of a finite dissipative layer similar to a wave passing through the air-glass-air as well as  a wave passing from a non-dissipative layer into a dissipative one (similar to the passage of light from air to water).

 The dissipation predictably results in reducing the soliton amplitude/velocity, but some new effects occur in the case of finite barrier on the soliton path: after the wave leaves the dissipative barrier it retains a soliton form, yet a reflection wave arises as small and quasi-harmonic oscillations (a breather). The breather spreads faster than the soliton as moves through the barrier.  \vspace{1mm}

\end{abstract}

\maketitle

\section{Introduction}

 The behavior of solutions to the KdV - Burgers equation is a subject of various recent research, \cite{key-2}--\cite{key-3}; the present paper is a continuation of \cite{key-1}.
 The aim is to study the behavior of the soliton  that, while moving in nondissipative medium encounters a barrier (finite or infinite) with finite dissipation. The modelling included the case of a finite dissipative layer similar to a wave passing through the air$\rightarrow$glass$\rightarrow$air as well as  a wave passing from a nondissipative layer into a dissipative one (similar to the passage of light from air to water).

 New results include a numerical model of the wave's behavior for different types of the media non-homogeneity. The dissipation predictably results in reducing the soliton amplitude, but some new effects occur in the case of finite  barrier on the soliton path: after the wave leaves the dissipative barrier it retains, on the whole, a soliton form yet a reflection wave arises as small and quasi-harmonic oscillations (a breather). This breather spreads as the soliton is moving through the barrier, and the breather moves faster than the soliton in the opposite direction. 
 
 For the modelling we used the Maple  \emph{PDETools} packet.

 The generalized KdV-Burgers equation considered here is of the form
 \begin{equation}\label{01}
    u_t(x,t)=\gamma(x)u_{xx}(x,t)- 2u(x,t)u_x(x,t)+ u_{xxx}(x,t).
    \end{equation}
 It is the simplest model for the medium which is both viscous and dispersive. The viscosity dampens oscillations except for stationary (or travelling wave) solutions.

 Note that $\gamma(x)\equiv 0$  corresponds to the KdV equations whose travelling waves solutions are solitons and $\gamma(x)\equiv\mathrm{const}>0$ corresponds to the KdV-Burgers equation  whose travelling waves solutions are shock waves. Tentatively
 for $\gamma(x)\neq 0$ the situation is similar to the geometric optics: as a ray enters  water from the air, one can observe the reflected wave and a decay of the transient wave.

 We consider the layered media which consist of layers with both dispersion an dissipation and layers  without dissipation. In the latter case the waves are described by the KdV equation, while  in the former --- by the Kdv-Burgers one. A soliton solution of the KdV equation, meeting a layer with dissipation, transforms somewhat similarly to a ray of light in the air crossing a semi-\-transparent plate. Thus  we consider three possibilities for $\gamma(x)$.

\begin{enumerate}
  \item $\gamma(x)=\alpha(1-\theta(x))$ is  the Heaviside step function(the two-layer case);
\item $\gamma(x)=\alpha(\theta(x-\beta)-\theta(x+\beta))$ is a $\Pi$-form density of viscosity (the three-layer case);
  \item $\gamma(x)=\alpha\cosh^{-2}2(\beta x)$ is a function with (numerically) compact support (the three-layer case).
\end{enumerate}

 We use the following initial value --- boundary problem (IVBP) for the KdV-Burgers equation on $\mathbb{R}$:
 \begin{equation}\label{08}
u(x,0) =6a^2\cosh(a(x+s)), \; u(\pm \infty,t) =0,\; u_x(\pm \infty,t) =0.
\end{equation}

 For numerical computations we use $ x\in[a,b]$ for appropriately large $a,\;b$ instead of $\mathbb{R}$.

\section{Soliton in 2-layer medium}

This case models a passage from non-dissipative half-space to a dissipative one. We take $\gamma(x)=0.5(1-\theta(x))$ as a  dissipation distribution function to present a single boundary separating these half-spaces.

Expect each solution  to behave as the one of the KdV at the right half-space   and as solution of KdV-B at the left one.

\begin{figure}[h]
\begin{minipage}{13.2pc}
\includegraphics[width=13.2pc]{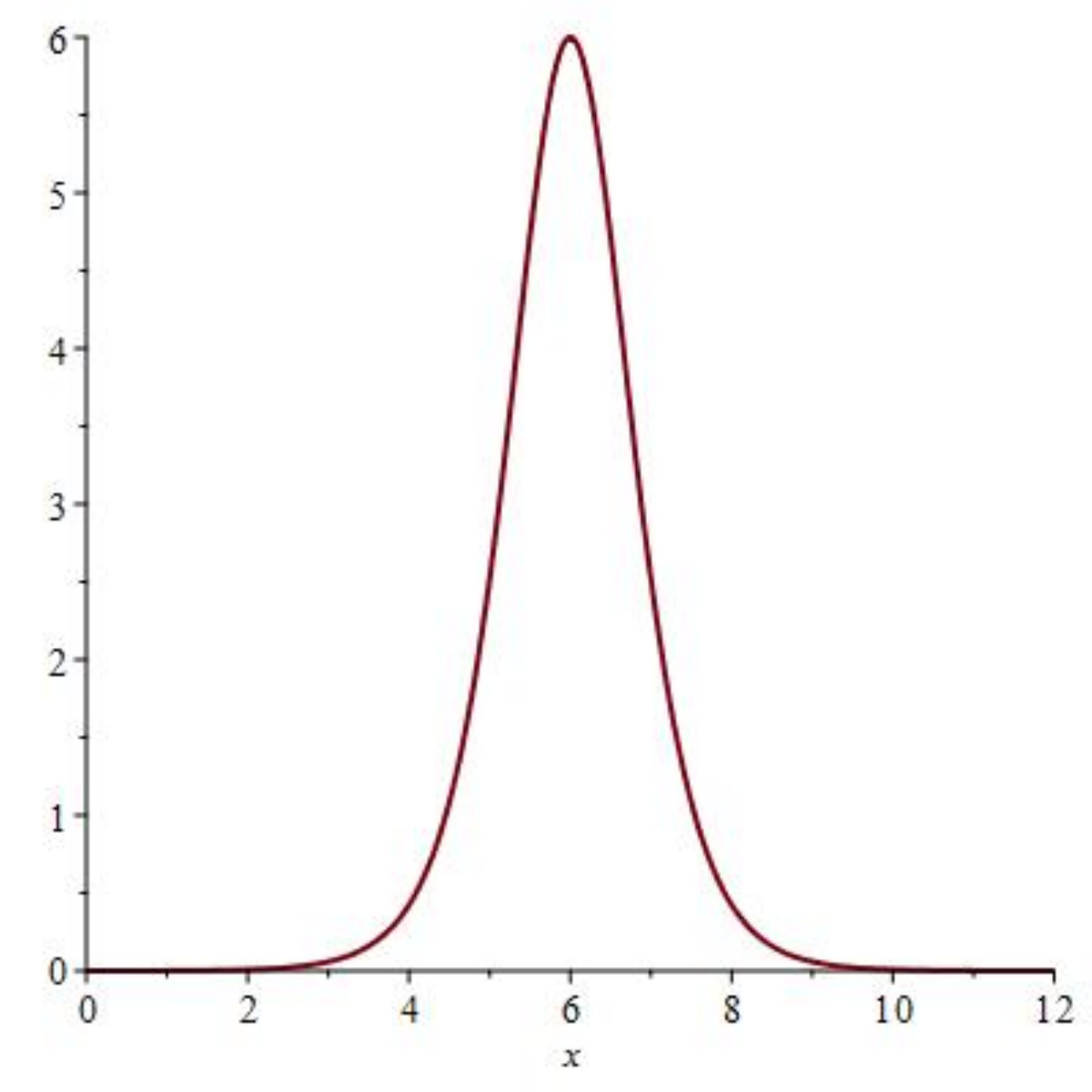}
\end{minipage}
\begin{minipage}{13.2pc}
\includegraphics[width=13.2pc]{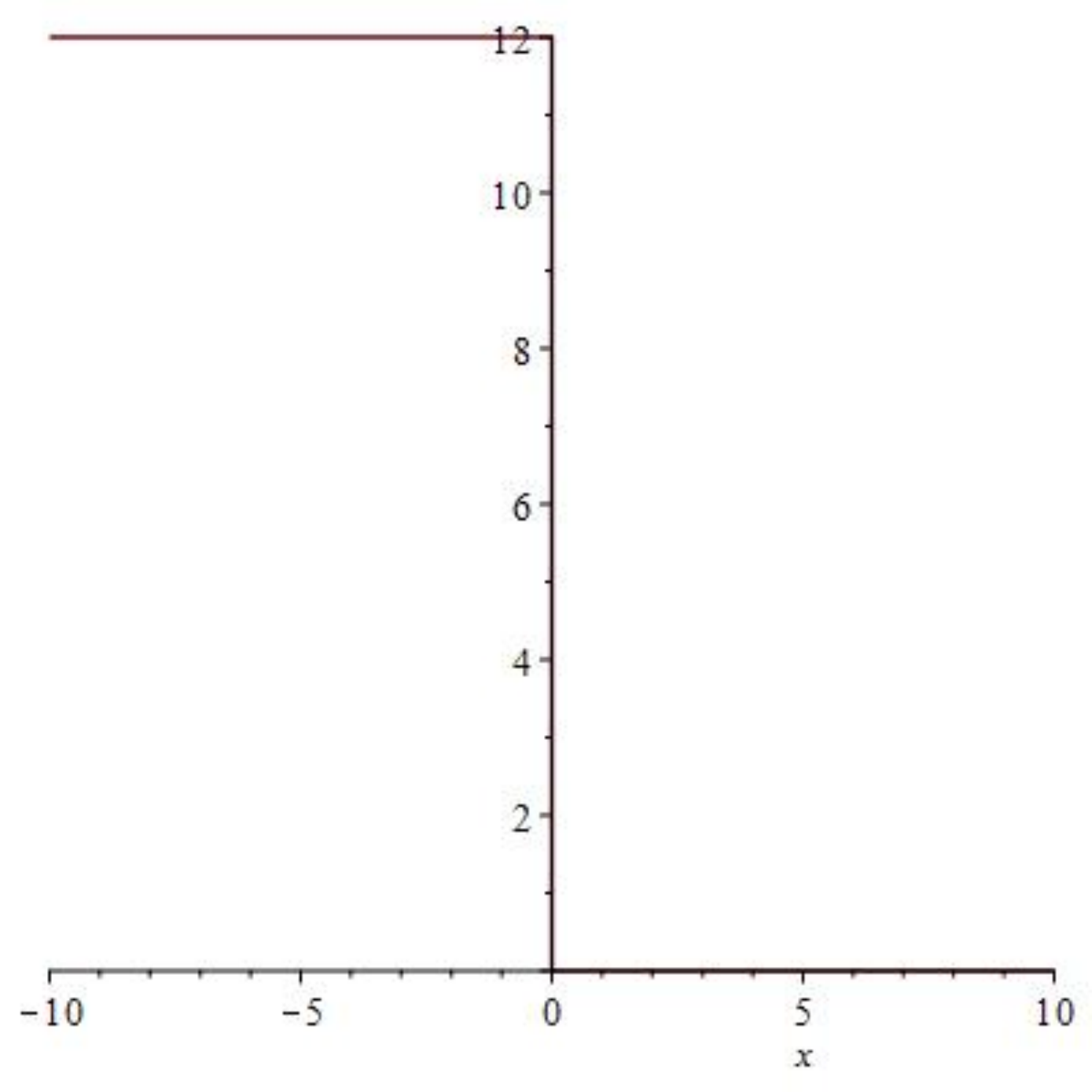}
\end{minipage}
\caption{\protect\small\textsl{\textbf{Left}:} Soliton $u(x,t,a,s)=6a^2 \cosh^{-2}(4a^3t+a(x+s))
$. Velocity $-4a^2=-4$, initial shift $s=-6$. \protect\newline\textsl{\textbf{Right}:} Step-like obstacle $\gamma(x)=\alpha(1-\theta(x)),\; \alpha=6 $}\label{ST1}
\end{figure}

\begin{figure}[h]
 \begin{minipage}{13.2pc}
\includegraphics[width=13.2pc]{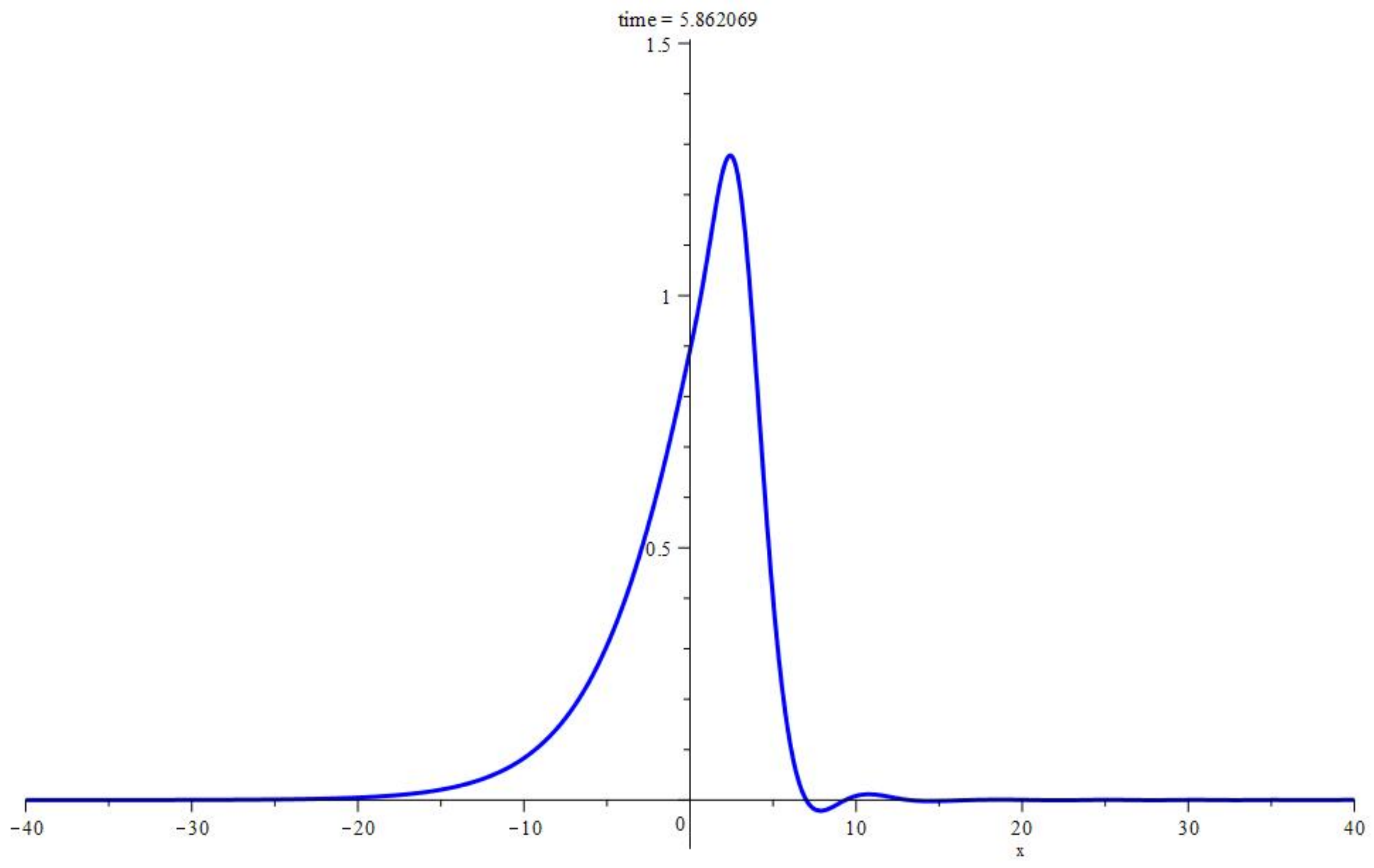}
\end{minipage}
 \begin{minipage}{13.2pc}
\includegraphics[width=13.2pc]{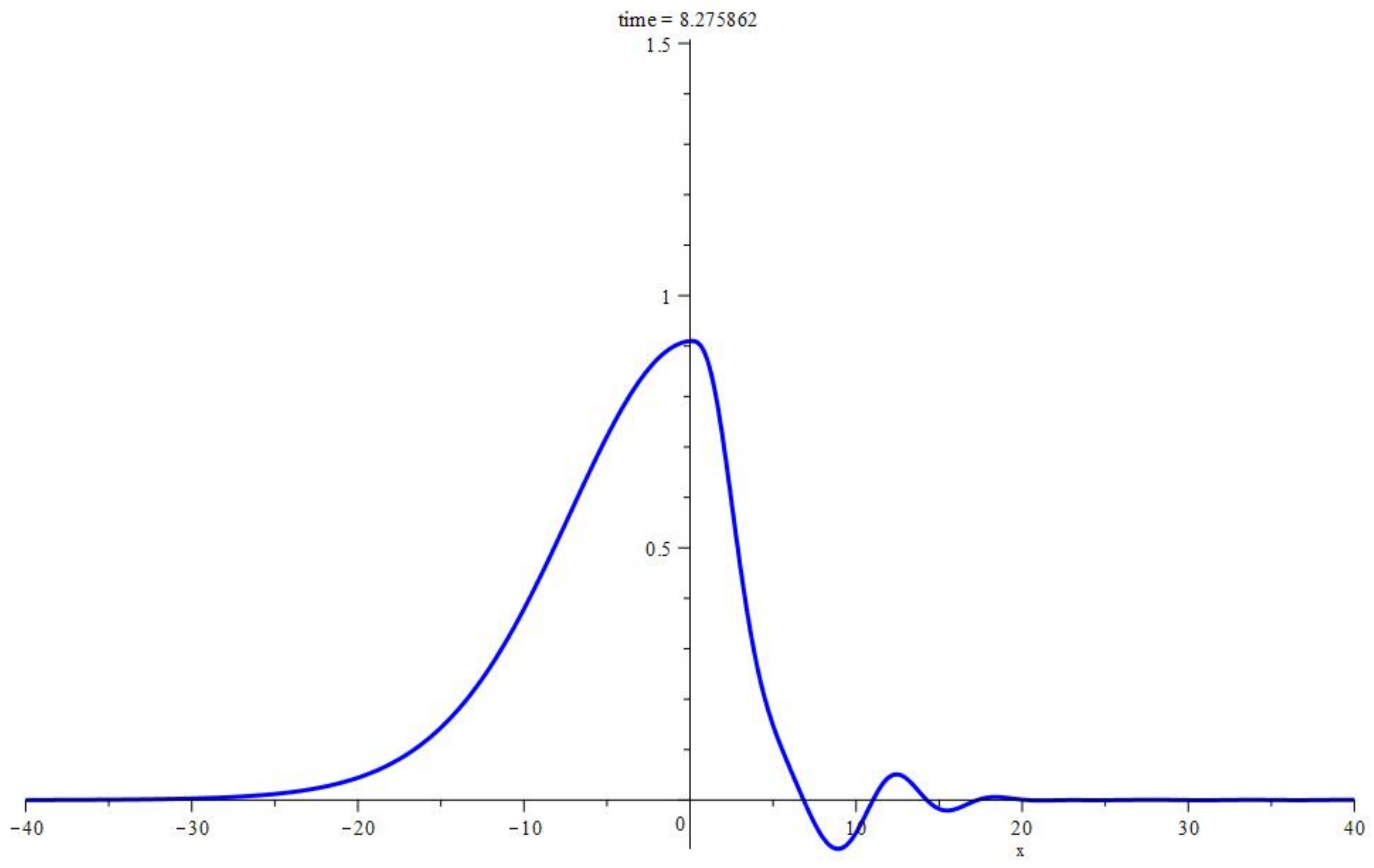}
\end{minipage}
\caption{ Soliton ($a=0.5$) passing a step-like obstacle ($\alpha=6$), \textsl{\textbf{Left}:} $t=6$; \textsl{\textbf{Right}:} $t=8$}\label{ST2}
\end{figure}

\begin{figure}[h]
  \begin{minipage}{13.2pc}
\includegraphics[width=13.2pc]{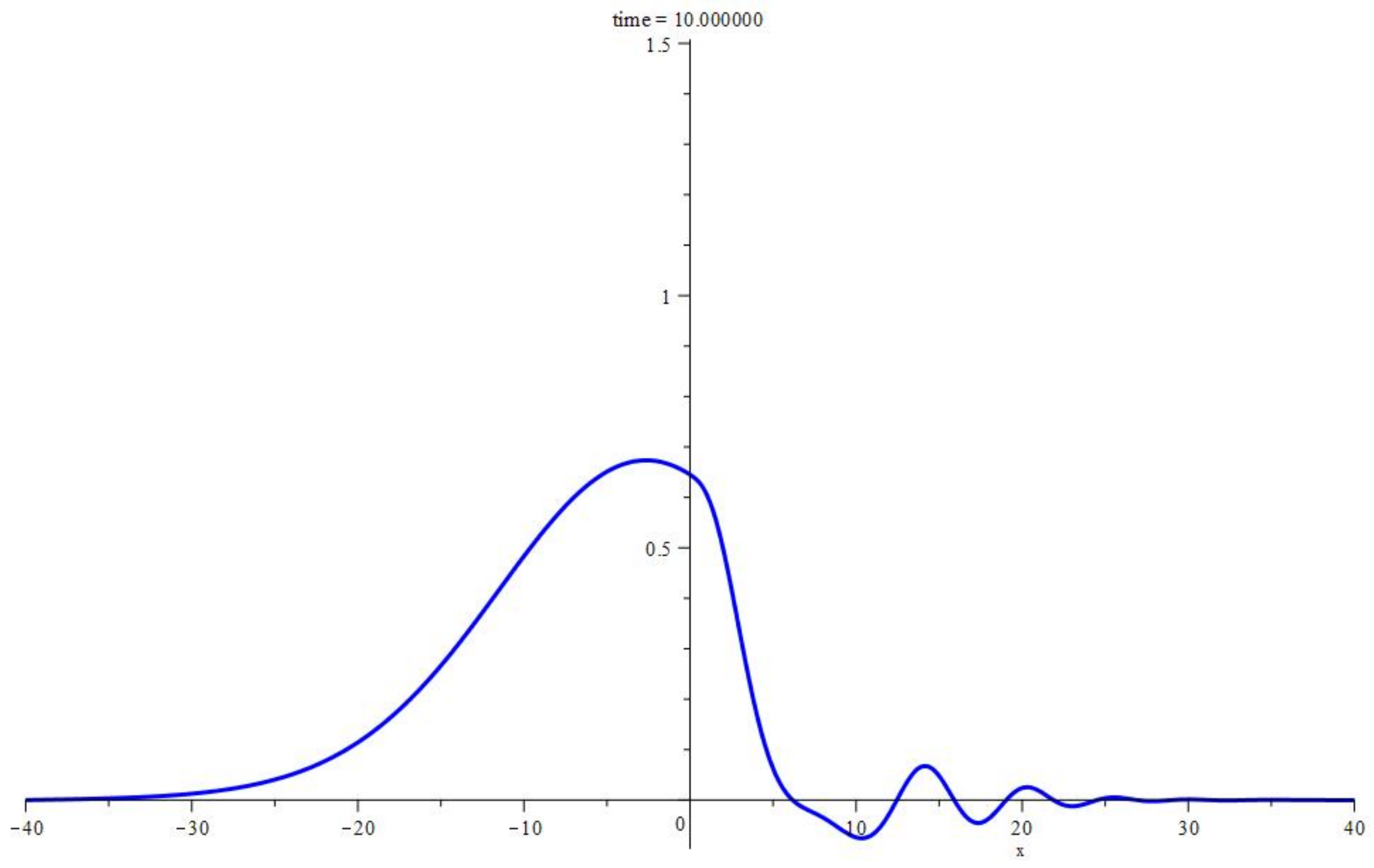}
\end{minipage}
 \begin{minipage}{13.2pc}
\includegraphics[width=13.2pc]{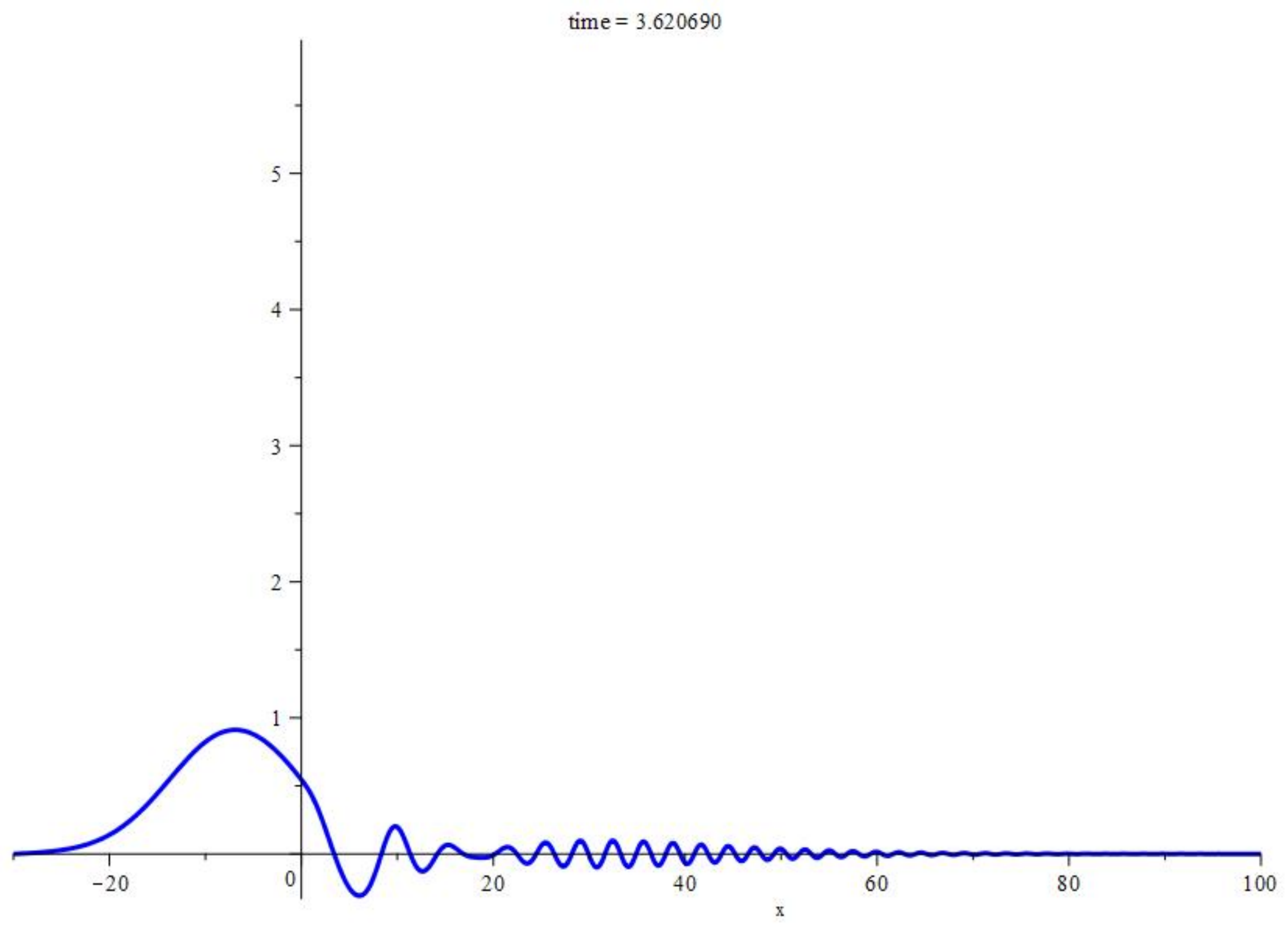}
\end{minipage}
\caption{\textsl{\textbf{Left}:} Soliton ($a=0.5$) passing a step-like obstacle ($\alpha=6$), $t=10$. \protect\newline\textsl{\textbf{Right}:} Soliton ($a=1$) passing a step-like obstacle ($\alpha=6$), $t=3.6$}\label{ST3}
\end{figure}

The process of transition from the soliton solution of the KdV to the correspondent solution of the KdV-B proceeds as expected. The transient wave in a dissipative media becomes a solitary shock which loses speed and decays to become nonexistent at $t\rightarrow +\infty$; and a reflected wave is seen in the non-dissipative half-space. It is similar to the passage of light from air into water; see figures \ref{ST1}--\ref{ST3}.

\section{$\Pi$ - type obstacle}

This case models a passage from non-dissipative half-space to another  one passing through a dissipative layer. We take $\gamma(x)=\alpha(\theta(x-\beta)-\theta(x+\beta))$ as a $\Pi$-form density of viscosity (the three-layer case) distribution function to present the layer separating these half-spaces.

Our experiments show that the initial soliton  behaves as the one of the KdV at the right half-space and as a diminished soliton  at the left one.

\begin{figure}[h]
\begin{minipage}{13.2pc}
\includegraphics[width=13.2pc]{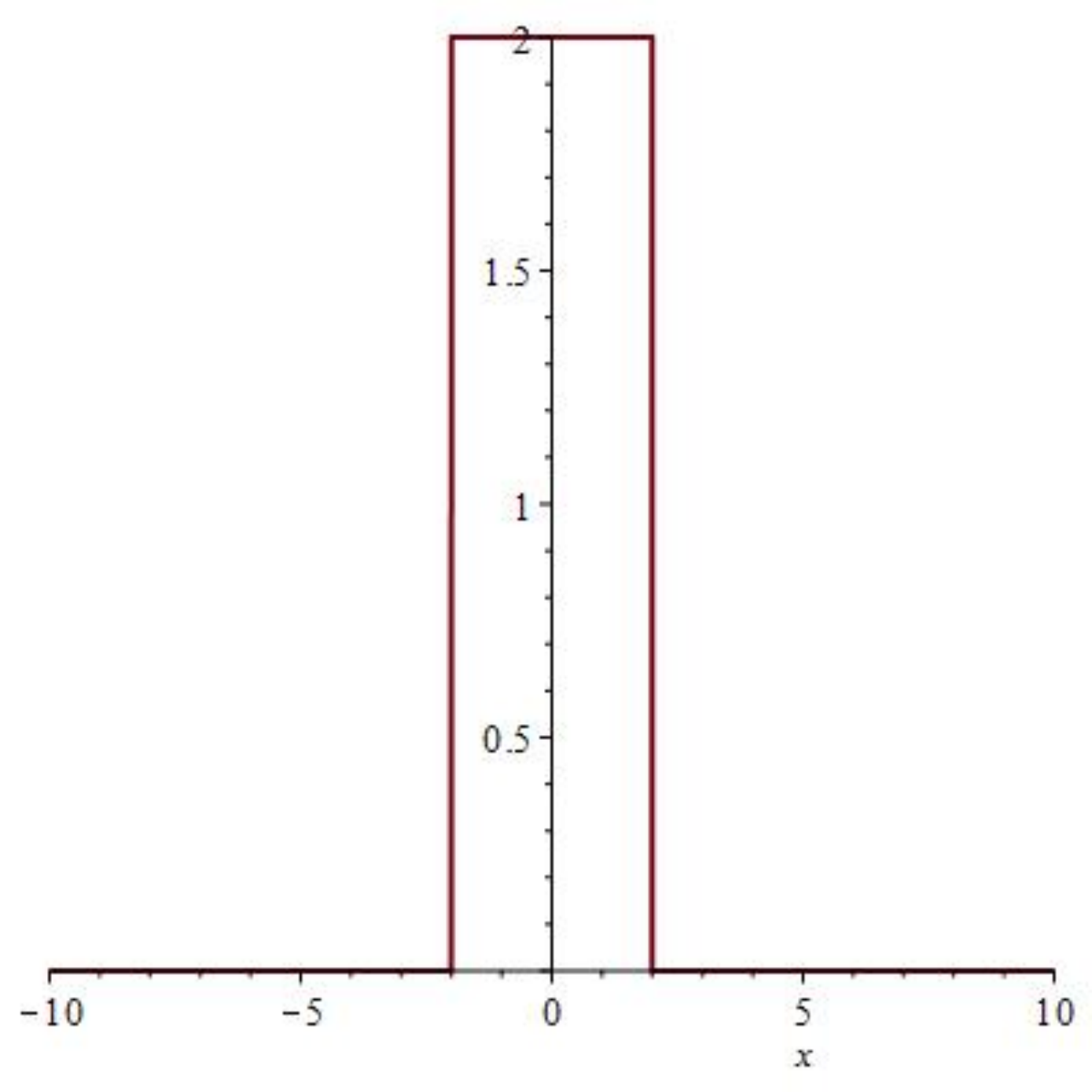}
\end{minipage}
\begin{minipage}{13.2pc}
\includegraphics[width=13.2pc]{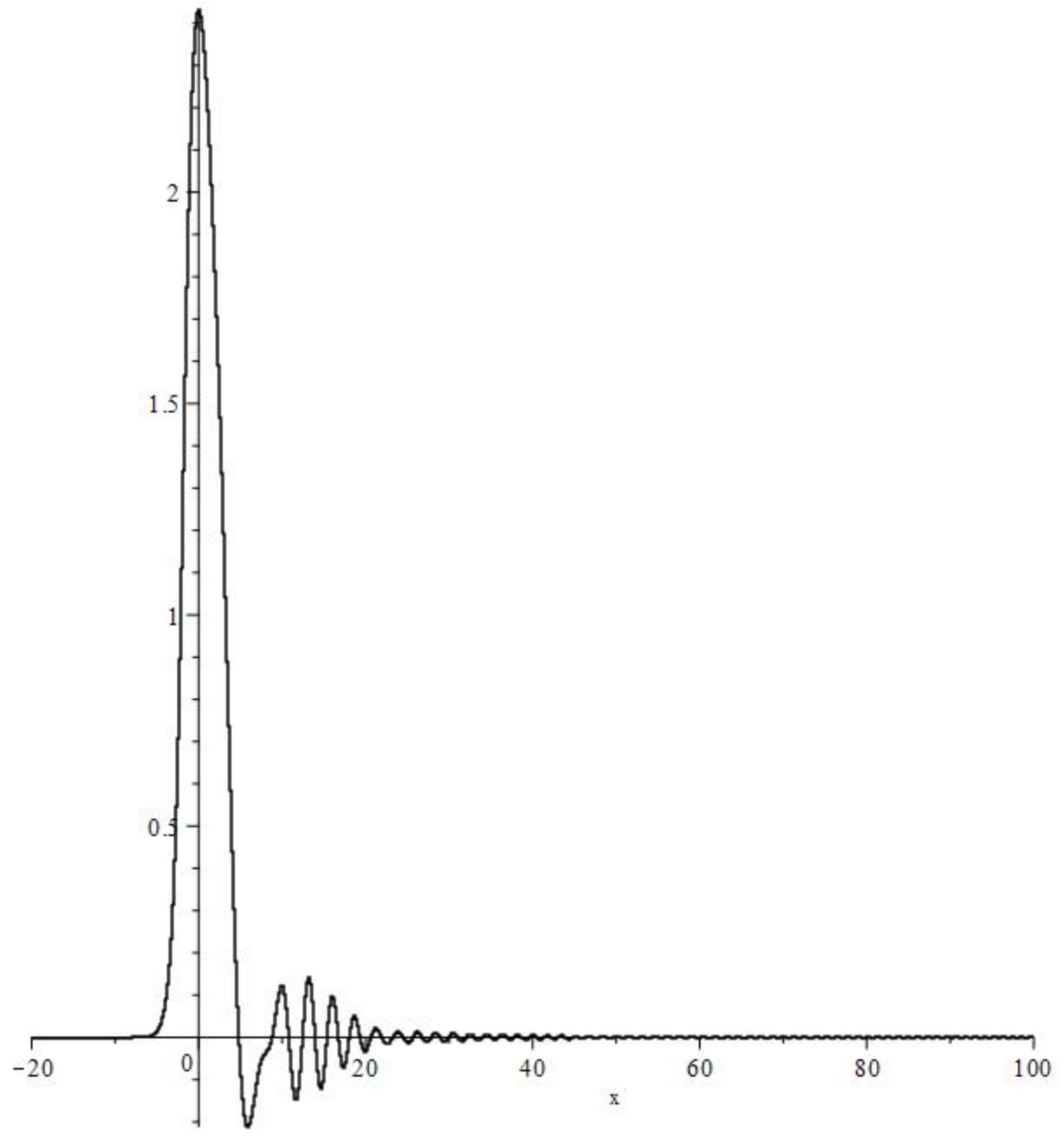}
\end{minipage}
\caption{\textsl{\textbf{Left}:} $\Pi$ - type viscose layer; $\alpha=1$, $\beta=2$. \protect\newline\textsl{\textbf{Right}:}  Soliton ($a=1$)  passing a thin viscose $\Pi$-type  layer ($\alpha=2.5$, $\beta=2$); $t=3$}\label{P1}
\end{figure}

\begin{figure}[h]
\begin{minipage}{13.2pc}
\includegraphics[width=13.2pc]{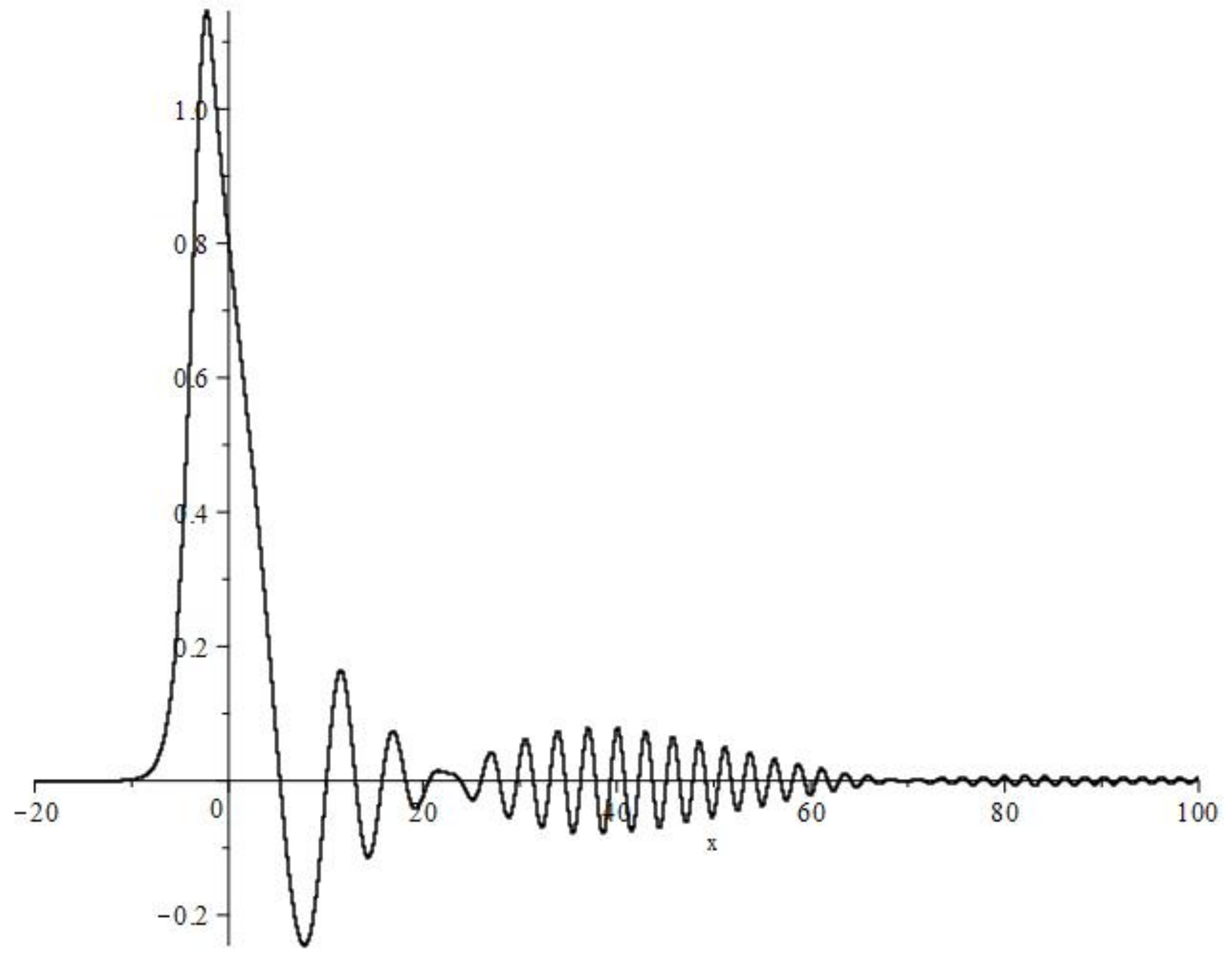}
\end{minipage}
\begin{minipage}{13.2pc}
\includegraphics[width=13.2pc]{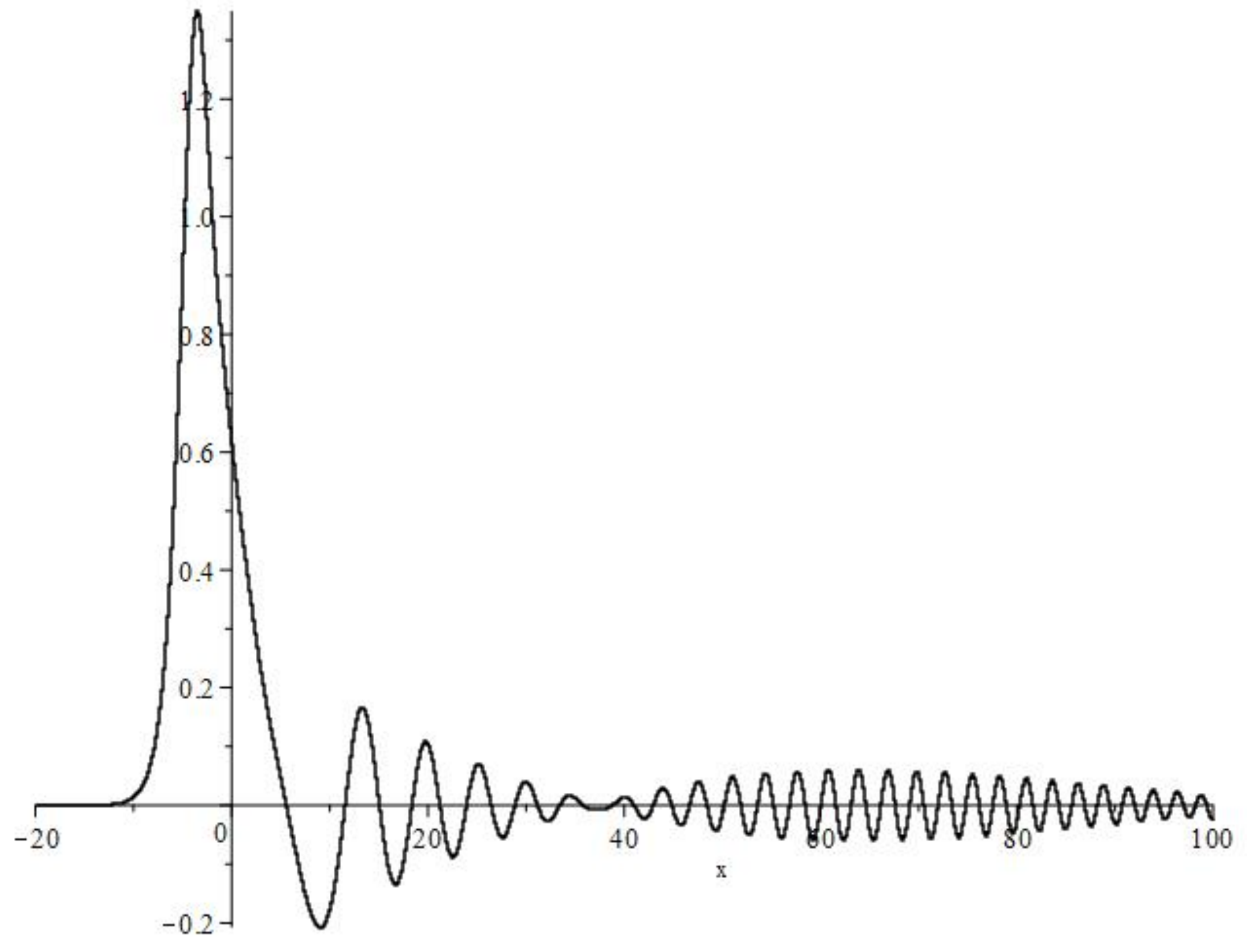}
\end{minipage}\caption{ Soliton ($a=1$)  passing a thin viscose $\Pi$-type  layer ($\alpha=2.5$, $\beta=2$). \textsl{\textbf{Left}:}   $t=5$ \textsl{\textbf{Right}:} $t=10$ }\label{P2}
\end{figure}

\begin{figure}[h]
\begin{minipage}{13.2pc}
\includegraphics[width=13.2pc]{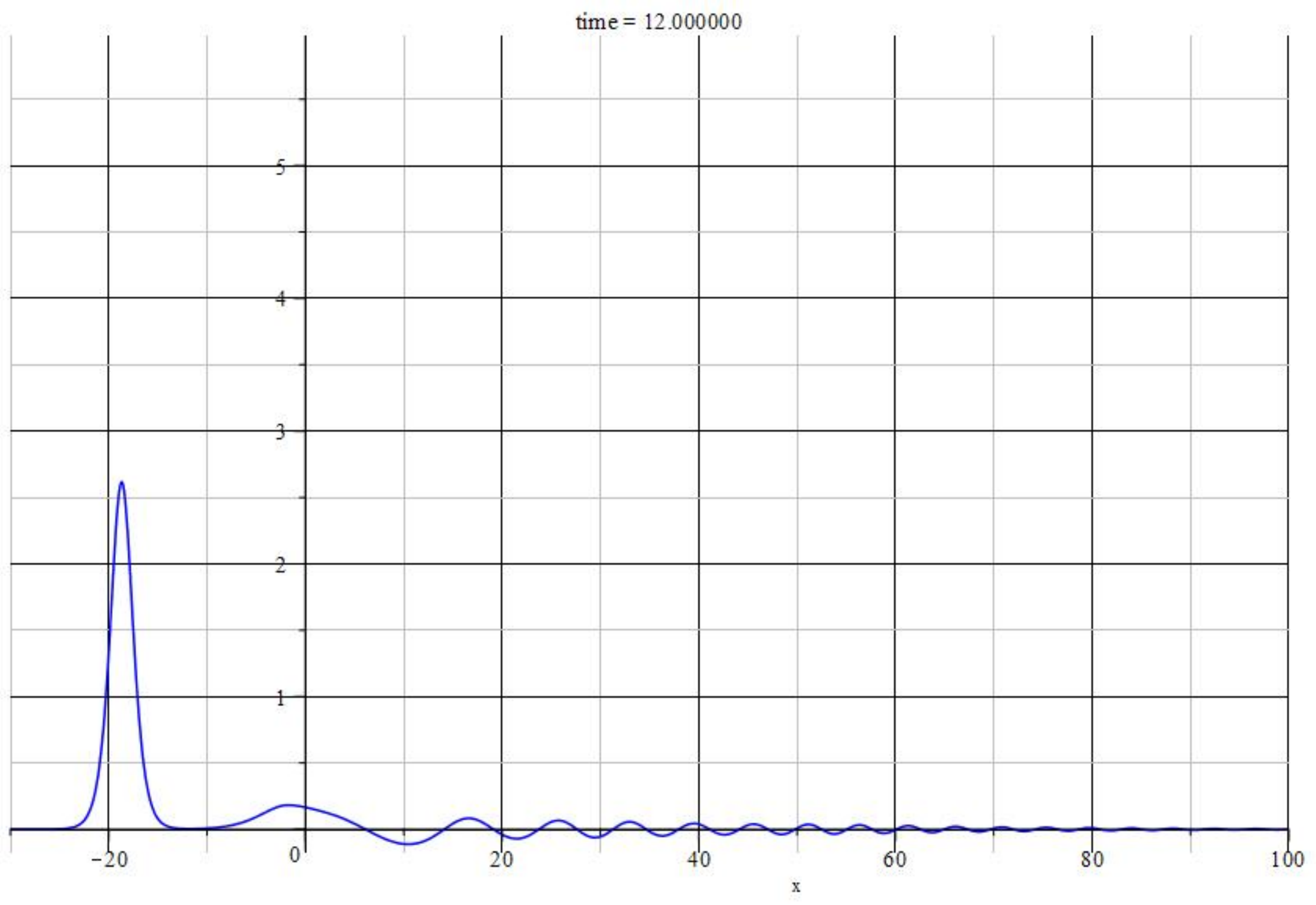}
\end{minipage}
\begin{minipage}{13.2pc}
\includegraphics[width=13.2pc]{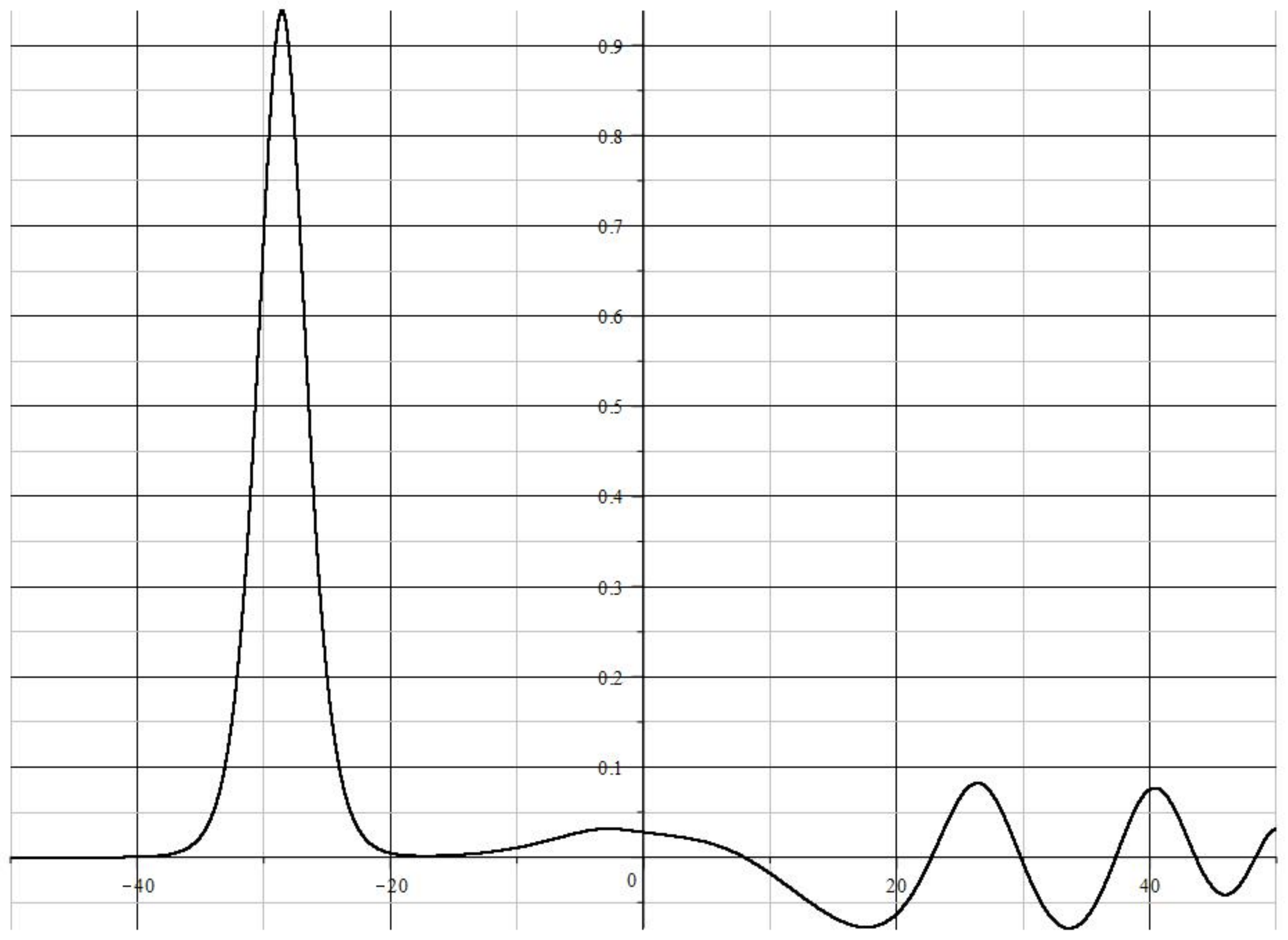}
\end{minipage}\caption{\textsl{\textbf{Left}:} Soliton ($a=1$) passing a $\Pi$ - type obstacle,  $t=12$;  \textsl{\textbf{Right}:} $t=50$.}\label{refraction}
\end{figure}

The process of transition  is natural enough. The transient wave in a dissipative media looses energy and speed to become a lesser and slower soliton at the left non-dissipative half-space; and a reflected wave is seen at the right half-space as it is shown at  figures \ref{P1}--\ref{refraction}.

The size of the reflected breather is connected, in particular, to the properties of the barrier $\alpha$ and $\beta$. So it may be of a practical use: for instance, measuring it one can judge whether the layer connecting two details is uniform at different points.

Since the refraction coefficient $k=\frac{V_1}{V_2}=\frac{4a_1^2}{4a_2^2}$ equals the ratio of amplitudes $\frac{6a_1^2}{6a_2^2}$, on figure \ref{refraction} (right) we see that $k\approx \frac{6}{0.95}\approx 6.3$.

\section{Soliton-like obstacle}

\begin{figure}[h]
\begin{minipage}{13.2pc}
\includegraphics[width=13.2pc]{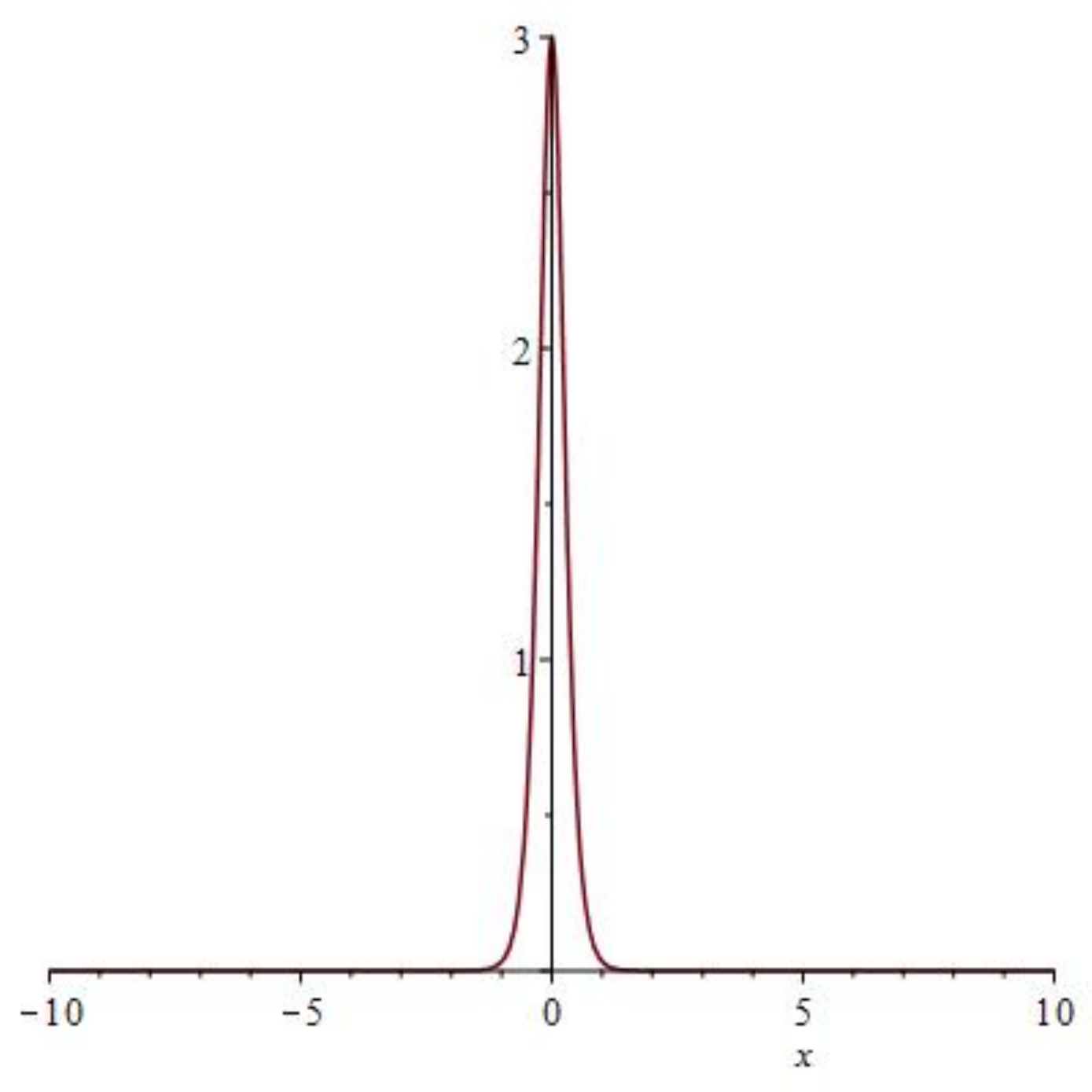}
\end{minipage}
\begin{minipage}{13.2pc}
\includegraphics[width=13.2pc]{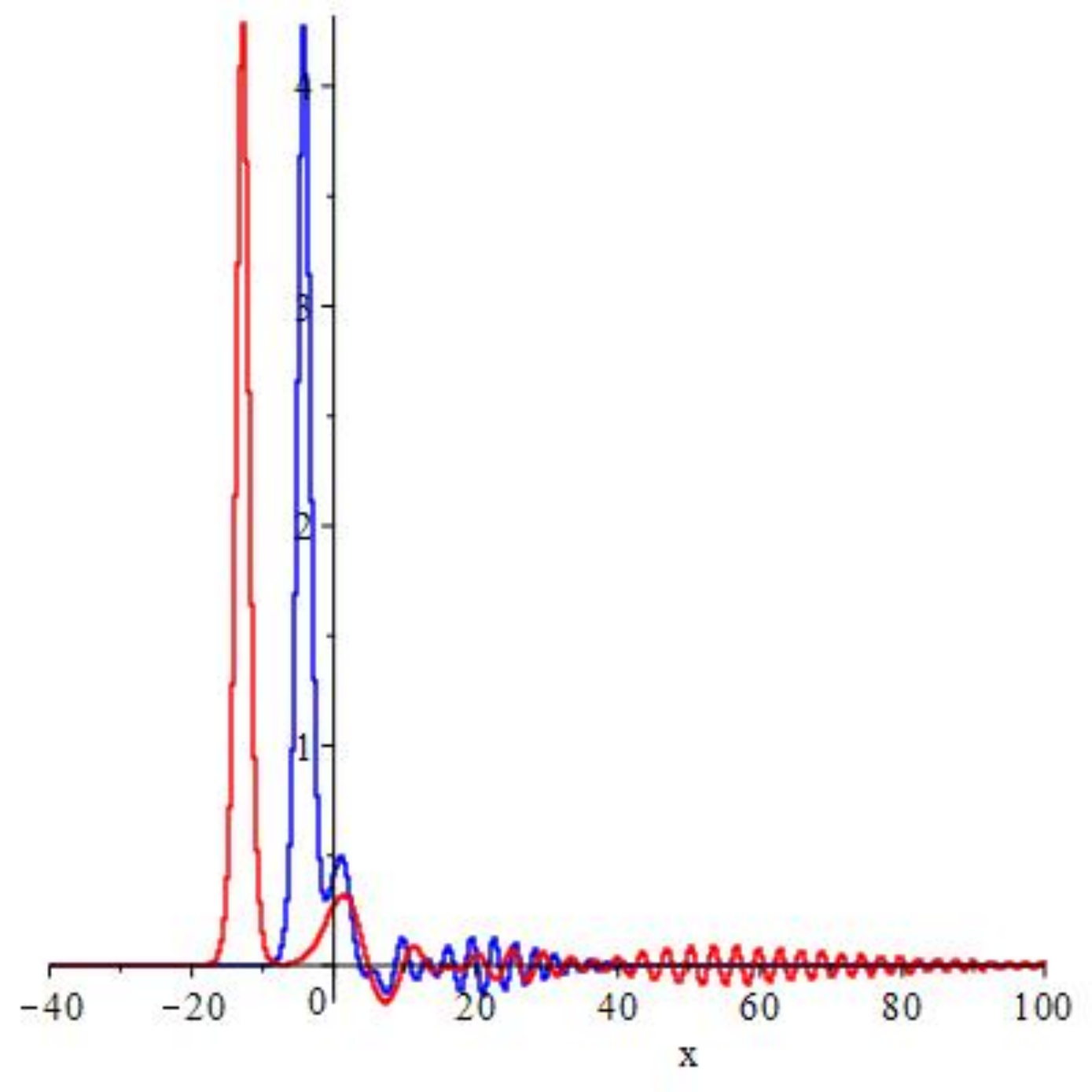}
\end{minipage}\caption{\textsl{\textbf{Left}:} Soliton-like obstacle, $\gamma(x)=3\cosh^{-2}(3x)$. \protect\newline\textsl{\textbf{Right}:} Soliton ($a=0.5$) passing a  thin viscose layer $\gamma(x)=3\cosh^{-2}(3x)$, $t=3$ (blue), $t=6$ (red) }\label{SL1}
\end{figure}

\begin{figure}[h]
\begin{minipage}{13.2pc}
\includegraphics[width=13.2pc]{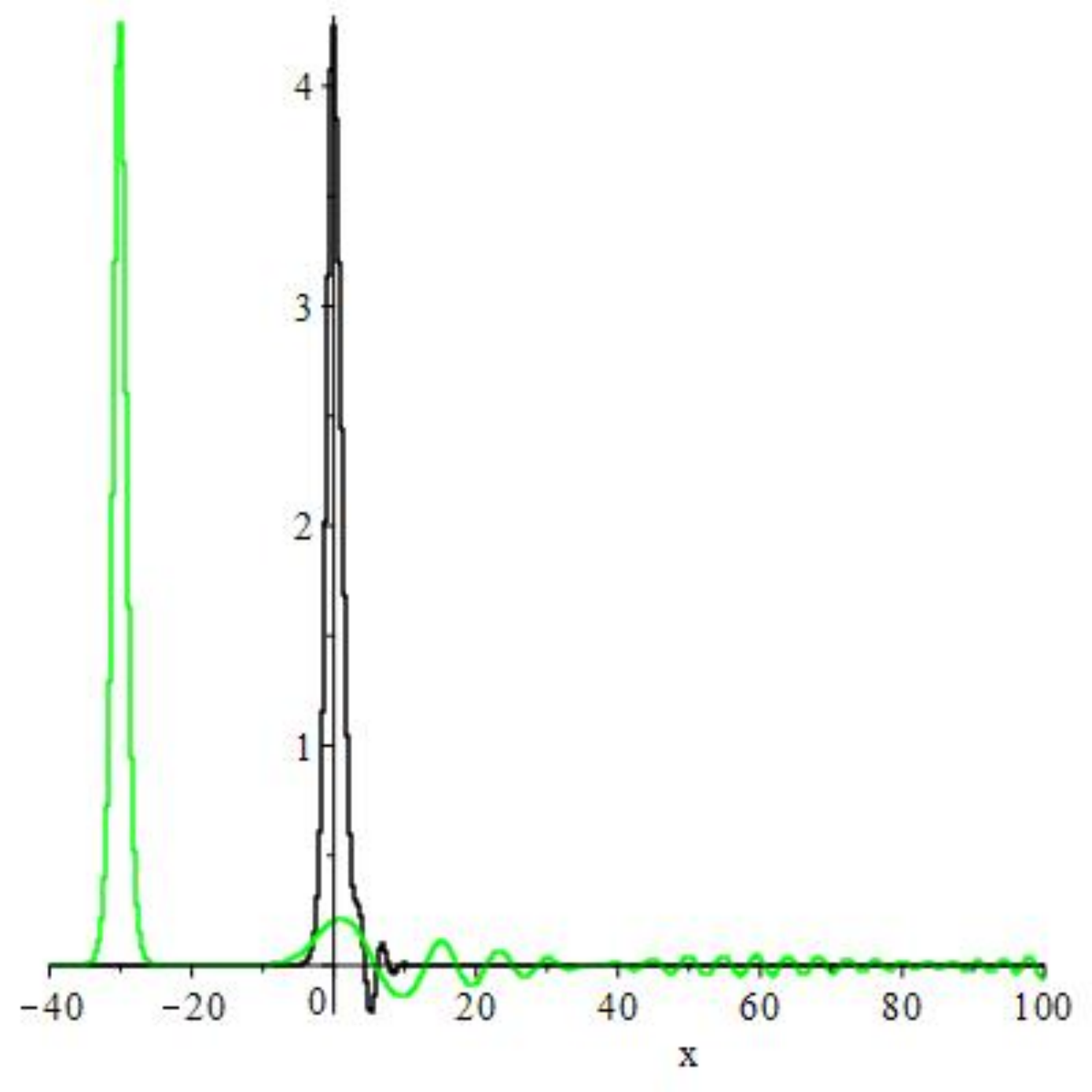}
\end{minipage}
\begin{minipage}{13.2pc}
\includegraphics[width=13.2pc]{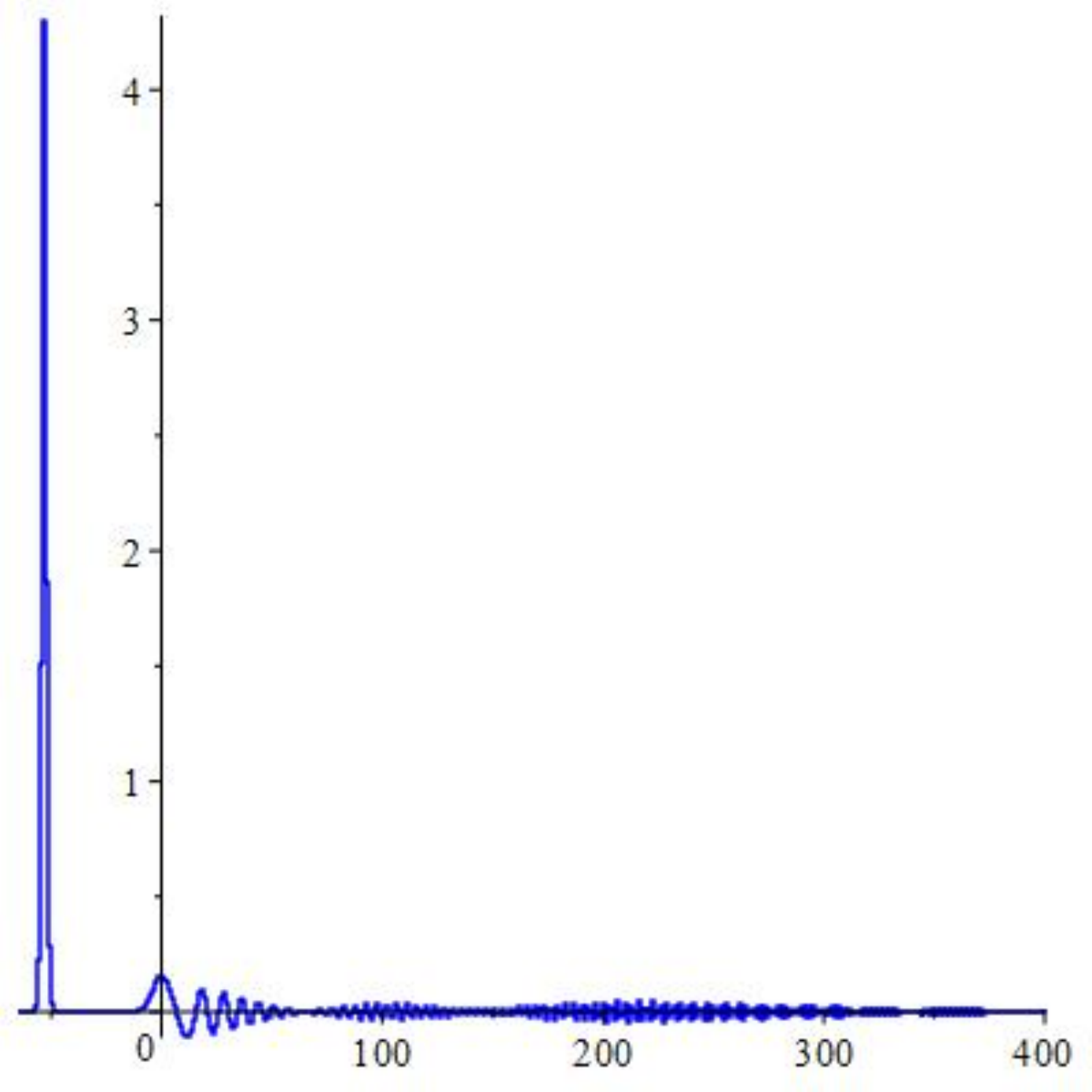}
\end{minipage}\caption{\textsl{\textbf{Left}:} Soliton ($a=0.5$) passing a thin viscose layer $\gamma(x)=3\cosh^{-2}(3x)$, $t=1.5$ (black), $t=12$ (green). \protect\newline\textsl{\textbf{Right}:} Soliton ($a=1$)  passing a thin viscose layer $3\cosh^{-2}(3x)$, $t=20$. The refraction coefficient $k\approx\frac{6}{4.3}\approx 1.4$}\label{SL2}
\end{figure}

Since the barrier $\gamma(x)=\delta\cosh^{-2}(\delta x)$ has a numerically compact support,  the transition process differs only slightly from that for the $\Pi$-type obstacle as it is illustrated by figures \ref{SL1}--\ref{SL2}.

\section{Other Travelling Wave Solutions}

The KdV equation (for $\gamma=0$) possesses the travelling wave solutions (TWS) of the form $ -6a^2\tanh(a(Vt+x)+s)^2+4a^2+(1/2)V.$

The soliton proper is such a TWS that tends to zero at $t\rightarrow\pm\infty$; thus $V=4a^2$; and it moves to the left.

\begin{figure}[h]
 \begin{minipage}{13.2pc}
\includegraphics[width=13.2pc]{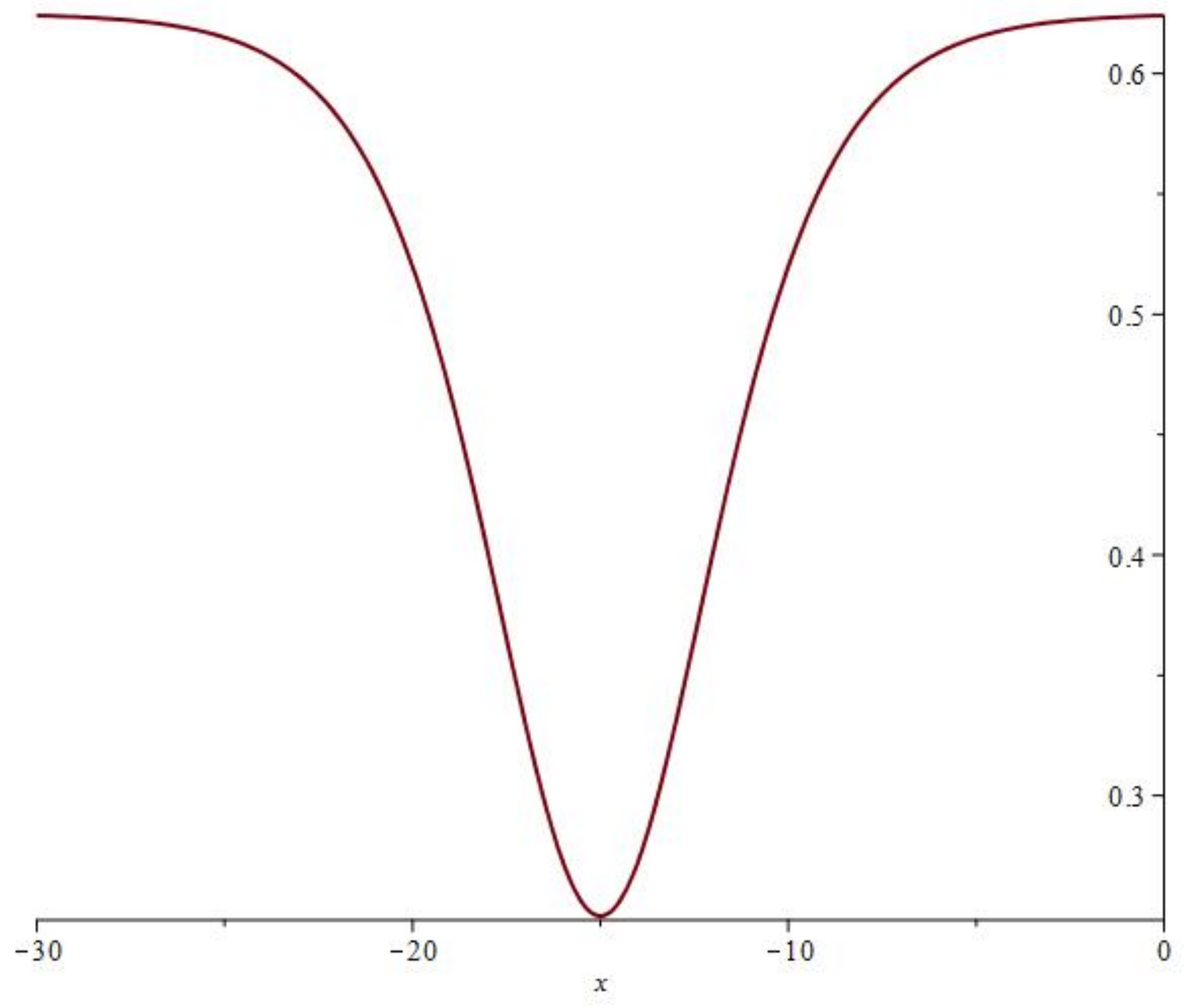}
\end{minipage}
\begin{minipage}{13.2pc}
\includegraphics[width=13.2pc]{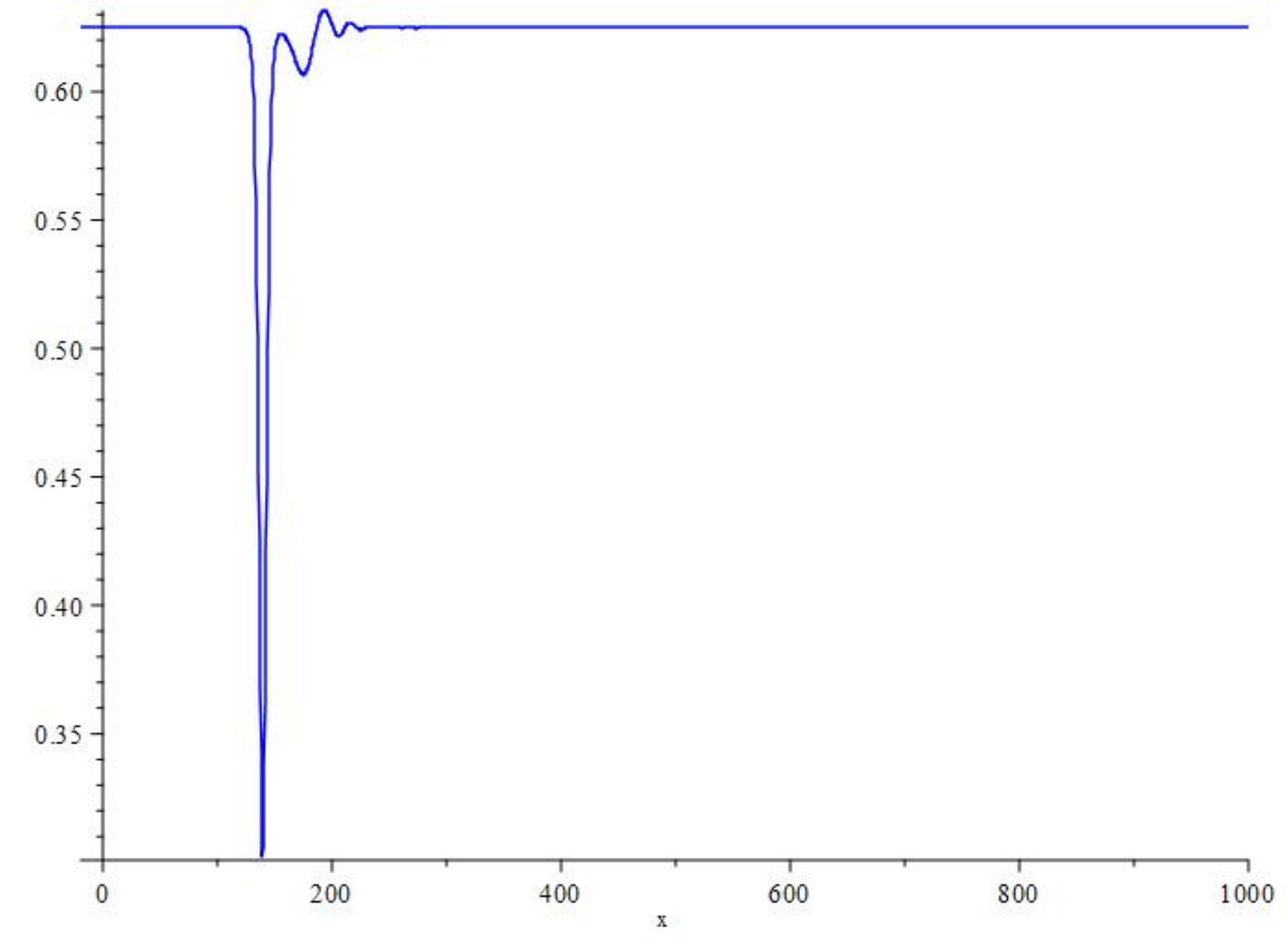}
\end{minipage}
\caption{\textsl{\textbf{Left}:} TWS: $V = -1., s = 3.75, a = 0.25$, moving to the right, at $t=0$. \protect\newline\textsl{\textbf{Right}:} TWS at $t=550$; $V = -1., s = 3.75, a = 0.25$}\label{TWS1}
\end{figure}

Take $V = -1., s = 3.75, a = 0.25$. The corresponding graph is presented
on figure \ref{TWS1}, left. The barrier is
$\gamma(x)=3\cosh^{-2}(3x)$.

\begin{center}
\begin{figure}[h]
\begin{minipage}{13.2pc}
\includegraphics[width=13.2pc]{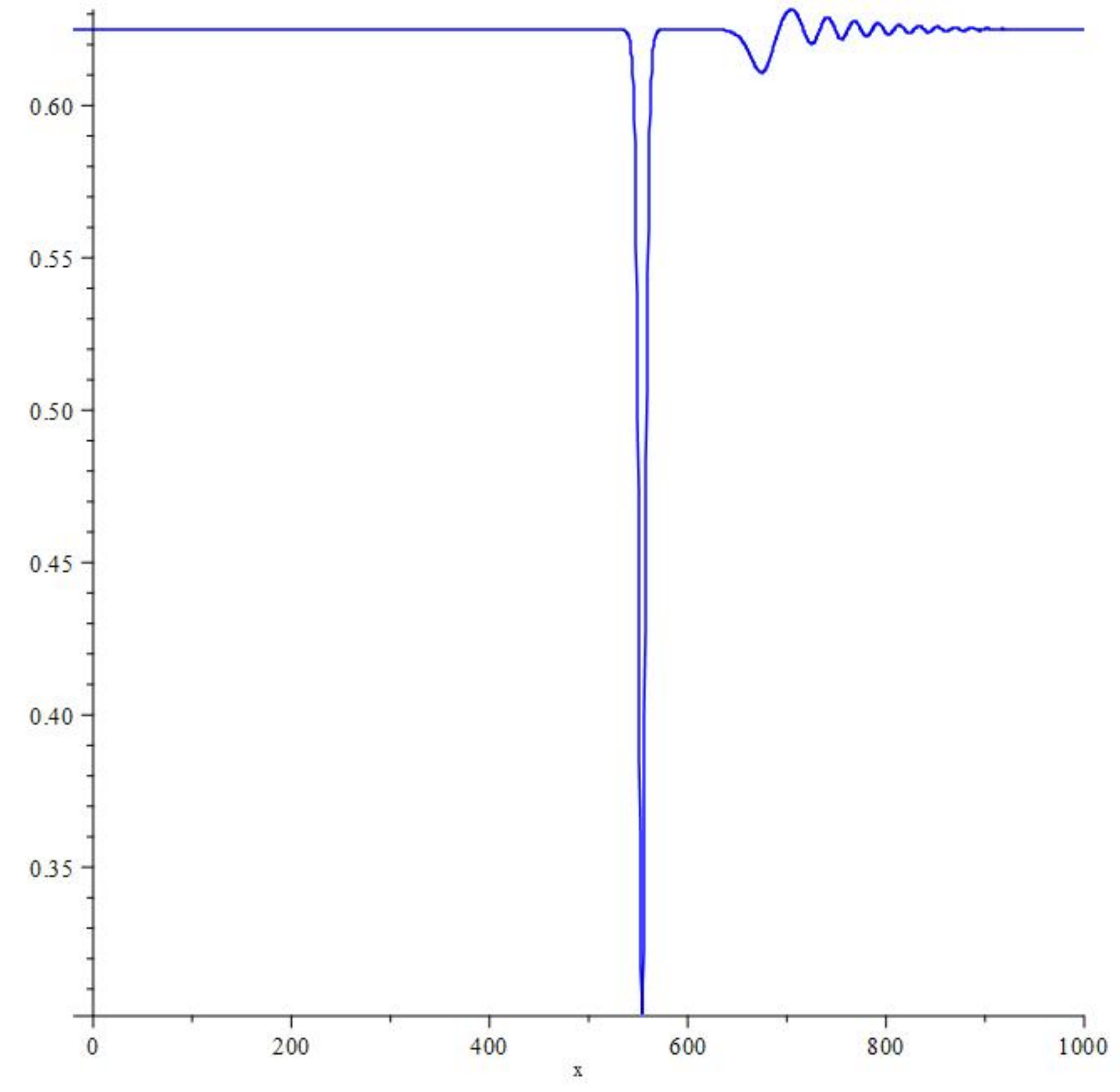}
\end{minipage}
\caption{TWS at $t=600$; $V = -1., s = 3.75, a = 0.25$}\label{TWS2}
\end{figure}
\end{center}

In is case both TWS and the breather move to the right and the breather moves ahead of the TWS;
the gap between  the TWS and the preceding breather  is widening over the time, see figures \ref{TWS1}--\ref{TWS2}.

\section{Conclusion and numeric considerations}

The results may be of a practical use. For once, the form of the reflected wave may be used to estimate the thickness and/or the density of the viscous barrier. A refraction may also be predicted.

The figures in this paper were generated numerically using Maple PDETools package. The  mode of operation uses the default Euler method, which is a centered implicit scheme, and  can be used to find solutions to PDEs that are first order in time, and arbitrary order in space, with no mixed partial derivatives. This implicit scheme is unconditionally stable for many problems (though this may need to be checked).

 Yet note that shocks and points of derivative's discontinuity is intrinsic for the model considered. Thus the standard procedures used with the default parameters may easily loose stability, which  leads to multi-oscillations and a general loss of precision. This problem was dealt with mainly by adapting  the \emph{spacestep} and/or the \emph{timestep} parameters.

 Qualitative estimations of the refraction coefficient, based on the relative decay of the KdV selected conservation laws will be published elsewhere.

\end{document}